\documentclass{mn2e}

\usepackage{graphicx}
\usepackage{color}

\def\today{\ifcase\month\or
 January\or February\or March\or April\or May\or June\or
 July\or August\or September\or October\or November\or
 December\fi\space\number\day, \number\year}

\def\todmy{\number\day\space\ifcase\month\or
 January\or February\or March\or April\or May\or June\or
 July\or August\or September\or October\or November\or
 December\fi\space\number\year}

\newcommand{\bdisp} {\begin{displaymath}}
\newcommand{\edisp} {\end{displaymath}}
\newcommand{\beqn} {\begin{equation}}
\newcommand{\eeqn} {\end{equation}}
\newcommand{\beqr} {\begin{array}}
\newcommand{\eeqr} {\end{array}}

\newcommand{\tal}{\it et al. \rm}
\newcommand{\AAA}{{A\&A}} 
\newcommand{\AAS}{{A\&AS}}
\newcommand{\ApJ}{{ApJ}}

\newcommand{\ApJS}{{ApJS}}
\newcommand{\AJ}{{AJ}}
\newcommand{\ApSS}{{Ap\&SS}}
\newcommand{\AR}{{ARA\&A}}

\newcommand{\MN}{{MNRAS}}

\newcommand{\PASJ}{{PASJ}}

\title{On the nature of bulges in general and of box/peanut bulges
  in particular. Input from $N$-body simulations.}  
\author[E. Athanassoula]
       {E. Athanassoula\\
Observatoire de Marseille, 
2 Place Le Verrier, 
F-13248 Marseille Cedex 4, France \\
}

\date{Accepted .
      Received ;
      }

\pagerange{\pageref{firstpage}--\pageref{lastpage}}
\pubyear{2004}

\begin{document}

\maketitle

\label{firstpage} 
\begin{abstract}
Objects designated as bulges in disc galaxies do not form a
homogeneous class. I 
distinguish three types. The classical bulges, whose
properties are similar to those of ellipticals and which form by
collapse or merging. Boxy and peanut bulges, which are seen in near edge-on
galaxies and which are in fact just a part of the bar seen edge-on. Finally
disc-like bulges, which result from the inflow of (mainly) gas to the
center-most 
parts, and subsequent star formation. I make a detailed comparison of the
properties of boxy and peanut bulges with those of $N$-body bars seen edge-on
and answer previously voiced objections about the links between
the two. I also present and analyse simulations where a boxy /peanut
feature is present at the same time as a classical spheroidal bulge
and compare them with observations. Finally, I propose a nomenclature
that can help distinguish between the three types of bulges and avoid
considerable confusion.
\end{abstract}


\section{Introduction}
\label{sec:intro}
\indent

Before studying any given class of objects, it is necessary to define
it, in order to select its members and exclude intruders. This first
step has not proven to be straightforward in the case of bulges in
disc galaxies and two definitions have been used so far.

According to the Webster's New Twentieth Century Dictionary of the English
Language (1963), a bulge is `the protuberant or more convex portion
of a thing; a part that swells out'. This led to the first definition
of a bulge : a smooth light distribution that
swells out of the central part of a disc viewed edge-on. The word
`smooth' stresses 
that the bulge should be constituted mainly of old stars, with little,
if any, star formation and dust. This definition is used
standardly in all morphological work. It can, and has, been used also 
in photometric studies. Then the bulge is defined as the central
region where the ellipticity of 
the isophotes is lower than that of the disc (e.g. Kent 1986;
Andredakis, Peletier \& Balcells 1995). 

A second definition stems from the analysis of radial photometric
profiles. Here the bulge is identified as the extra light in the
central part of the 
disc, above the exponential profile fitting the remaining
(non-central) part. In earlier papers this component was fitted with an
$r^{1/4}$ law, while more recent ones use its generalisation to an
$r^{1/n}$ law, commonly known as S\'ersic's law (S\'ersic 1968). This
definition was adopted by Carollo, Ferguson \& Wyse (1999) and has the
advantage of being applicable to disc galaxies independent of their
inclination. It has also the advantage of leading
to quantitative results about the mass distribution. Nevertheless, it
has the disadvantage of assigning to the bulge any extra central
luminosity of the disc, above the inwardly extrapolated
exponential, independent of its origin. I will show in this paper some
cases where this can lead to confusion.

These two definitions, however, are too
general and apply to more than one type of objects, with quite
different physical properties and formation histories.   
Indeed bulges are not a homogeneous
class of objects. A very large number of studies of the observational
properties of bulges have stressed their similarity with
ellipticals. This includes their structure, their photometry as well
as their kinematics (see e.g. reviews by
Illingworth 1983; Wyse, Gilmore \& Franx 1997; and references
therein). Yet, as more objects were considered and as the  
quality of the observations increased, discrepancies became
clear. Illingworth (1983) argued that some bulges have shapes which
are never seen in ellipticals, namely peanuts or strongly boxed
shapes. He also stressed the different kinematic properties of a
handful of bulges (Kormendy \& Illingworth 1982; Davies \& Illingworth
1983) from those of ellipticals, particularly with respect to their
$V/\sigma$ values. Kormendy (1993) argued that a number of  
central components, which by the above definitions 
would be considered as bulges, have many properties bringing them
closer to discs than to spheroids. This includes their shape, their
kinematics and the
fact that they have inner spiral or ring structures, bright
star-forming knots and dust lanes
(e.g. Carollo, Stiavelli \& Mack 1998; Carollo \& Stiavelli
1998). 
The debate on the spheroid versus disc nature of bulges has
continued in a large number of papers (see reviews and collection of
papers in e.g. Dejonghe \& Habing 1993; Wyse, Gilmore \& Franx 1997;
Carollo, Ferguson \& Wyse 1999; Balcells 2003; Kormendy \& Kennicutt
2004). 

$N$-body simulations have repeatedly shown
that bars seen side-on\footnote{I call side-on the edge-on view in
  which the line-of-sight is along the bar minor axis and end-on the
  edge-on view in which the line of sight is along the bar major
  axis.}
have a boxy/peanut shape 
(e.g. Combes \& Sanders 1981; Combes \tal 1990, hereafter CDFP90; Raha
\tal 1991, hereafter RSJK91; Athanassoula \& Misiriotis 2002,
hereafter AM02; Athanassoula 2002, hereafter A02; Athanassoula 2003,
hereafter A03; O'Neill \& Dubinski 2003; Martinez-Valpuesta \&
Shlosman 2004). These simulations argued for a link between bars and
peanut features and against the formation of peanuts from
interactions (Binney \& Petrou 1985, Rowley 1988), unless of course
the interactions excite or strengthen a 
bar component. A very important argument in favour of this link comes from
kinematical observations and their interpretation (Kuijken \&
Merrifield 1995; Merrifield \& Kuijken 1999; Bureau \& Freeman 1999;
Bureau \& Athanassoula 1999, 2005; Athanassoula \& Bureau 1999; Chung
\& Bureau 2004).  

Bulges, or more precisely bulge-to-disc ratios, are one of the three
criteria for classifying disc galaxies (e.g. Sandage 1961).
This makes it particularly crucial for us to understand what a
bulge really is. The fact that this class of objects is so
inhomogeneous has generated a number of misunderstandings and has 
hampered progress in the subject. Considerable confusion has also been
brought by the fact that the formation scenarios presented in a number of 
papers include a mixture of the seemingly appropriate buzz-words,
but are otherwise unclear as to the exact formation procedure they propose. It
is thus time to get back to 
basics and examine what a bulge really is, and disentangle the
different members of this class. This paper aims to contribute
to this goal.

In section~\ref{sec:3types} I distinguish three different types of
bulges. Section~\ref{sec:comparison} presents $N$-body results and
compares them with observations of boxy/peanut
bulges. It also summarises similar comparisons available in the
literature. Section~\ref{sec:objections} addresses the objections presented
so far to the argument that peanut/boxy bulges are just
bars seen edge-on. I propose a nomenclature which can help distinguish
between the various types of bulges and conclude briefly in
section~\ref{sec:last}.

\section{Three different types of bulges}
\label{sec:3types}

Schematically, one can recognise three different types of
bulges. I will here distinguish them via their formation histories and
then discuss the observational properties these lead to. This approach
is clearly limited by the fact that some scenarios have not been fully
worked out yet. The opposite approach, namely distinguishing
objects from their observational properties and then asking what
formation scenarios could lead to such properties, would also suffer
from the same limitation and would, furthermore, be more dependent on the
incompleteness of the observational picture. I have thus chosen here
the first approach. It should of course be kept in mind that the
types presented here are schematic and that more than one process may
have contributed to the formation of a given bulge. This will be
discussed further at the end of the section.

{\bf Classical bulges}. These are formed by gravitational collapse or
hierarchical merging of smaller objects and corresponding dissipative gas
processes. The formation process is generally fast and sometimes
externally driven. It occurs early on in the galaxy formation process,
before the present discs were formed. 

Several versions of this scenario have been elaborated in
simulations. In the simulations of Steinmetz \& M\"uller (1995) 
the bulge is formed during the first starburst triggered by the
collapse of small-scale density fluctuations and is composed mainly of
old metal-rich stars. A bulge forms before the disc also in the
simulations of Samland \& Gerhard (2003) and Sommer-Larsen \tal
(2003). In the simulation of 
Steinmetz \& Navarro (2002) a disc was formed which at $z \sim $ 3.3
merged with another galaxy of a similar size. This stirred the stars
from the disc component and concentrated them into a bulge
progenitor. It also triggered a burst of star formation that depleted
most of the gas of the initial discs. 

A further variant of this scenario involves the formation of a gaseous
proto-disc which forms clumps via gravitational instabilities. These
clumps spiral to the center by dynamical friction and merge to form a
central bulge. This 
was first proposed by Noguchi (1998, 1999) and fully worked out, in
more realistic multi-phase simulations, by Immeli \tal (2004a,
2004b). If the disc containing the clumps is viewed edge-on before the
clumps merge, it resembles the chain galaxies, first observed by
Cowie, Hu \& Songaila (1995), who tentatively assigned them to
redshifts between 0.5 and 3. Subsequent observations by Elmegreen \tal
(2004a, 2004b) with the Hubble Space telescope gave a similar, but
somewhat narrower, redshift range of 0.5 to 2. The time-scale for the
clumps to spiral in is rather short, of the order of two disc rotation
times. The star formation rate in the clumps and particularly in the
bulge is very high, leading to a large overabundance of
$\alpha$-elements of the bulge stars.  

Bulges formed in this way should have several similarities to elliptical 
galaxies, including their photometric radial profiles, their
kinematics and their stellar populations (e.g. Davies \tal 1983; Franx
1993; Wyse, Gilmore \& Franx 1997; and references therein). Thus they should
be composed of predominantly old stars, they should have predominantly  
ellipsoidal shapes and should have near-$r^{1/4}$ projected density
profiles. A typical object in this category is e.g. the Sombrero 
galaxy (NGC 4594). Although a relatively large mass
(compared to the disc), as in NGC 4594, is tell 
tale of a classical bulge, this is not a prerequisite. 

The prevalent view is that formation of classical bulges should
happen early on in the galaxy 
formation process, before the present discs were actually
formed. Nevertheless, it may be possible to build a bulge by accretion at
a much later stage, after the disc has grown, as suggested by the 
simulations of e.g. Pfenniger (1993), Athanassoula (1999) Aguerri, Balcells \&
Peletier (2001) or Fu, Huang \& Deng (2003). Aguerri \tal consider the
merging of a small elliptical with a disc galaxy which, 
before the merging, has a small bulge with an exponential radial
density profile. After the merging the bulge has increased its mass,
while its radial density profile has steepened considerably, reaching an
$r^{1/4}$ if the satellite mass is equal to that of the
proto-bulge. Athanassoula (1999 and in prep.) considers a
similar event, but with an initially bulge-less disc. The merging can
occur after the thin disc has formed, provided the angle of the
equatorial plane of the disc galaxy with the orbital plane of the
satellite is not big. The stellar populations
of the bulge will be those of the small elliptical that formed it and
the time necessary for the small elliptical to reach the center
of the target will depend
drastically on its mass, massive ellipticals spiraling in faster than
less massive ones. 

{\bf Box/peanut bulges}. These objects are formed via the
natural evolution of barred galaxies. Bars form
spontaneously\footnote{Their formation may, in some cases, be helped
  by the gravitational interaction with a companion.} in
disc galaxies and then evolve at a slower rate. The time necessary for
the initial bar formation is longer for galaxies whose halo within the
inner few disc scale-lengths is relatively more massive
(Athanassoula \& Sellwood 1986; A02). Relevant time 
scales are of order of a few galaxy rotations. Somewhat after bar
formation some of the material in the bar acquires stronger vertical
motions and thus reaches larger distances from the equatorial
plane. These distances increase with time. Viewed edge-on, this gives a
characteristic box/peanut shape.   

Objects formed in this way should have observed morphological, photometrical and
kinematical properties that are the same as those of $N$-body bars
seen edge-on, as I will discuss in section~\ref{sec:comparison}. Since
they form by rearrangement of disc material,
they should be constituted of stellar populations that are similar
to those of the inner disc at radii comparable to those of the
box/peanut feature. Subsequent star formation in one, or both, of
these components can introduce some young stars. The age of the
bulk of the stars, however, can be considerably older than the age of the
boxy/peanut feature itself. The average size of these features
should be of the order of 1 to 3 disc scale-lengths, and can not
reach $D_{25}$. This formation scenario has been well worked out in a number of
papers describing relevant $N$-body simulations (e.g.  CDFP90; RSJK91;
AM02; A03; O'Neill \& Dubinski 2003), while the orbital structure
responsible for such boxy/peanut features has been studied in considerable
detail (Pfenniger 1984; Skokos, Patsis \& Athanassoula 2002a,b,
hereafter SPAa and 
SPAb; Patsis, Skokos \& Athanassoula 2002, hereafter PSA02; Patsis, Skokos \&
Athanassoula 2003, hereafter PSA03; and references therein). 

{\bf Disc-like bulges.} Contrary to the boxy/peanut bulges, the
formation scenario of  discy bulges is not fully worked out. But the general
picture is as follows : It is well known that gas will
concentrate to the inner parts of the disc under the influence of the
gravitational torque of a bar, thus forming an inner disc extending
roughly up to the (linear) inner Lindblad resonance, or forming a ring at such
radii 
(e.g. Athanassoula 1992; Wada \& Habe 1992, 1995; Friedli \& Benz
1993; Heller \& Shlosman 1994; Regan \& Teuben 2004). The extent of
this region is of the order of a kpc. When this 
disc/ring becomes sufficiently massive it will form stars, which
should be observable as a young population in the central part of
discs. Kormendy \& Kennicutt (2004) estimate that the star formation
rate density in this region is $0.1 - 1~M_{\odot} yr^{-1} kpc^{-2}$, i.e. 1
to 3 orders of magnitude higher than the 
star formation rate average over the whole disc. This will lead
naturally to the formation of a sizeable central disc. Unfortunately,
no detailed $N$-body simulations of this scenario, including star
formation and comparisons with observations of central discs, have
been published so far, although many aspects have been well studied
individually and much work is in progress. Note that disc-like bulges
can also be formed in $N$-body simulations with no gas, from inwards motions of
the disc material (see sect.~\ref{subsec:Nbody}). These disc-like
bulges, however, will be less massive than
those formed by combined stellar and gaseous processes.

The formation scenario of disc-like bulges could be, and in several cases
has been, evolved further. Namely, if the disc-like bulge is
sufficiently massive and  
concentrated, it could destroy partially or totally the bar that
formed it. Various authors have advocated the mixing of the material
that was initially in the bar with that of the young disc 
described above, to form a more vertically extended object. Yet
this scenario needs to be better understood. The efficiency of bar
dissolution mechanisms still remains 
uncertain, since simulations have given widely different
estimates of the necessary mass and of its central concentration
(Hasan, Pfenniger \& Norman 1993; Friedli 1994;
Norman, Sellwood \& Hasan 1996; Hozumi \& Hernquist 1998, 1999; Berentzen
\tal 1998; Bournaud \& Combes 2002; Shen \& Sellwood 2004;
Athanassoula, Dehnen \& Lambert 
2003 and in prep.). Furthermore, $N$-body simulations have
not studied in sufficient detail the changes in the vertical structure that
accompany this dissolution. Can the peanut or box survive the bar
weakening, or its dissolution?
More work is necessary on this and on several other aspects of
this scenario. Seen its weaknesses, I will not consider this second
part of the formation scenario of disc-like bulges further in this
paper and will constrain myself to the first part.

Bulges formed with this scenario can have properties attributed normally
to disc systems and can contain substructures found normally in
discs. They can contain a sizeable amount of gas, as well as stars 
younger than those formed with the two previous scenarios. They qualify as 
bulges by the second definition given in sect.~\ref{sec:intro}, but not
by the first one. 

Objects whose properties correspond to the above formation scenario
have indeed been identified, with the help of the second definition 
(Sect.~\ref{sec:intro}), i.e. with the help  of radial photometric
profiles. The substructures they harbour, mainly spirals, rings,
bright star forming knots, dust lanes and 
even bars, have been discussed e.g. in Kormendy (1993), Carollo,
Stiavelli \& Mack (1998) and Kormendy \& Kennicutt (2004).
Their radial photometric profiles are closer to exponential than to
$r^{1/4}$ (e.g. Courteau, de Jong \& Broeils 1996, 
Carollo, Stiavelli \& Mack 1998) and their
colours imply a younger age than that of $r^{1/4}$ bulges
(e.g. Carollo \tal 2001). Yet they can be observed also in the near
infrared. They are primarily found in late type disc galaxies
(e.g. Andredakis \tal 1995; Carollo \& Stiavelli 1998) as expected
since gas processes  can enhance their formation.
Their host galaxies often harbour a large-scale bar.
Prominent strong central peaks have indeed been observed in
radial photometric profiles obtained from 
cuts along the major axis of edge-on barred galaxies (see e.g. Fig. 1, 2 and
5 in L\"utticke, Dettmar \& Pohlen 2000). These features are only seen
in cuts which are on, or near the equatorial plane, which argues for
their disc-like geometry. 

In summary, the objects in this category 
are disc-like in shape and have many disc-like properties but
form strong excesses on radial photometric profiles, reminiscent
of those of bulges, i.e. they qualify as bulges by the second, but not
by the first definition. The link of these disc-like objects with
bulges is not novel, and has indeed been made in a number of studies (see
e.g. Kormendy \& Kennicutt 2004 and references therein).

Although the processes leading to either the second or the third type
of bulges are both bar driven, they are different. The
processes involved in the formation of boxy/peanut bulges invoke
vertical instabilities and are necessarily dissipationless. On the
other hand the one invoked for discy bulges involves radial
redistribution of material and 
can rely at least partly on the presence of a dissipational component. Moreover,
the two classes of objects thus formed are distinctly different. In
particular, one 
extends to radii comparable to (although, as we will see later, somewhat 
smaller than) that of the bar itself, while the other should be
constrained to considerably smaller regions, of the order of a
kpc. Their vertical extents are also very different. Their only common
point is that they are both 
made from disc material by bar-driven evolution, contrary to the first
type of bulges, which is in no way linked to a pre-existing bar. This
one and only similarity, however, is not sufficient for them to be considered
together, as a single class of objects.

The three types of bulges are very different but are not mutually
exclusive. Face-on strongly barred galaxies often harbour what
looks like classical bulges and should thus contain both a classical
and a box/peanut bulge. Furthermore, the strong central peaks of
the radial density profile in edge-on boxy/peanut systems
(e.g. L\"utticke \tal 2000), which
I have linked to disc-like or dissipative bulges in the
previous paragraphs, show that boxy/peanut bulges often coexist with
disc/dissipative ones.  

Samland \& Gerhard (2003) presented a simulation in which two types of
bulges form at different times. During the early formation stages
a classical bulge forms by dissipative collapse. This is followed by
the formation of a disc which is bar unstable and forms what the authors call a
bar-bulge. This has axial ratios roughly 3:1.4:1 and includes the old
bulge component formed in the early collapse. Thus the bulge in their
model contains
two stellar populations : and old population that formed during the
collapse phase and a younger bar population, the distinguishing feature
between the two being the [$\alpha$/Fe] ratio. Such bulges, containing
two stellar populations, have also been found observationally
(e.g. Prugniel \tal 2001) and stress the fact that the various types
of bulges can coexist in the same galaxy.

\section{Comparison of properties of bars in simulations viewed
  edge-on with observations of boxy/peanut features}
\label{sec:comparison}

$N$-body simulations have now reached sufficient resolution to allow a
number of detailed comparisons with observations. I will here briefly
discuss 
those concerned with the morphology, photometry and kinematics of 
boxy/peanut features. 

\subsection{$N$-body results}
\label{subsec:Nbody}

I have at my disposal more than 250 simulations describing bar
formation in disc galaxies. For about 200 of these I used initial conditions
similar to those described in A03, while roughly 50 used initial 
conditions as described by Kuijken \& Dubinski (1995). All simulations
have more than a million particles (2 $\times 10^5$ in the disc and
roughly $10^6$ in the halo), while a few of them have
considerably more, up to 5 millions. They were
run either on our GRAPE5 systems (Kawai \tal 2000) or with a
treecode on a PC (Dehnen 2000, 2002). More
information on how the simulations were run and the codes that were
used can be found in A03, while the system of computer units is as
described in AM02. I used a large number of
these simulations to study the observable properties of bars seen
edge-on. Thus, the results presented in this section are based on a
very large sample, although for practical reasons only a few
simulations are used for illustration.

Athanassoula (A02, A03) used both analytical calculations and
$N$-body simulations to show that it is the exchange of angular
momentum within the galaxy that drives the bar evolution and,
in particular, that determines its strength, pattern speed and
morphology. 
The reader is referred to these two papers for more information, as
well as for references on both the
analytical and numerical work on the subject.
Angular momentum is emitted from the bar region, mainly at the resonances,
and absorbed by the outer disc and by the halo, again mainly at the
resonances. Since the bar is a negative angular momentum
feature, it becomes stronger and slows down by losing angular momentum.
The more angular momentum it can shed, 
the stronger and slower it can become. Thus, if the disc is immersed in
a receptive (massive and cool) halo, it may be able to shed more of
its angular momentum and thus grow stronger than in the absence of
such a halo, or than in the presence of a rigid halo (A02; A03). 
Here I use the same notation as in AM02, A02 and A03. In
particular, 
models with haloes which are not too hot and which have important
contributions to the total 
rotation curve within the whole galaxy, including the innermost few disc
scale-lengths, are called MH.
If they also have a classical bulge, then they are termed
MHB. On the other hand,  
models which have a disc that dominates dynamically the inner region
are termed MD (or MDB if they also have a classical bulge). 

Peanuts form spontaneously in bar unstable 3D $N$-body simulations (e.g. 
CDFP90; RSJK91; AM02; A02; A03; O'Neill \& Dubinski 2003). 
Such features, however, do not form simultaneously with the bar, but
some time after 
it, as was already shown by CDFP90 and is confirmed by my
simulations. Thus the bar starts thin and then 
buckles out of the plane, initially asymmetrically, and finally 
tending towards symmetry with respect to the equatorial plane. This
evolution can be seen e.g. in the figures of  RSJK91, as well as in my
simulations. My 
simulations also show clearly that it is not the whole bar that
buckles, so that the outermost part of the bar stays thin and
planar.

Stronger peanuts generally grow in models with stronger bars. This
could already be  
inferred from the few examples in AM02, while a more quantitative
approach, based on a large number of simulations, will be presented
elsewhere. In particular, considering a sequence
of bars with varying axial ratio, 
I note that the strong, thin bars have X or strong peanut
features, while ovals have weak peanuts or boxy structures. Considering now a
sequence of bars with varying length, I note that the boxy/peanut
feature stands out less clearly in very short bars. Yet, at least in all 
simulations I checked, it was always possible to see it in cuts parallel to
the equatorial plane, or on isocontours.
  
\subsection{Morphology}

\begin{figure*}
\begin{center}
\setlength{\unitlength}{2cm} 
\includegraphics[scale=1.0,angle=0]{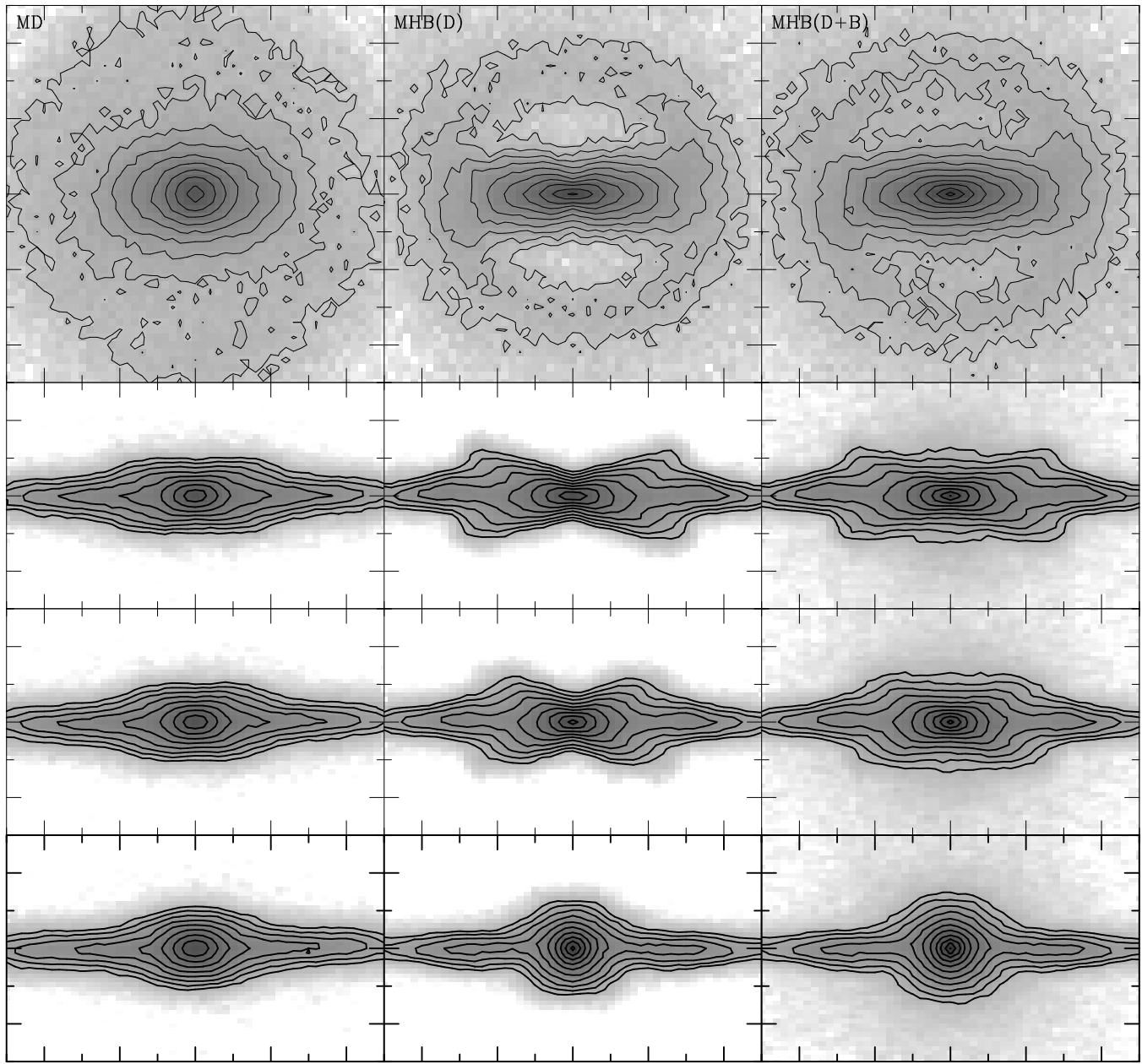}
\caption{Face-on (upper row), side-on (second row), $45^{\circ}$
  viewing angle (third row) and end-on 
  (lower row) views for two characteristic simulations. 
  The projected density is given by  
  grey-scale and also by isocontours (spaced logarithmically). The
  left column of panels illustrates a simulation of MD type, while the
  central and 
  right columns show a simulation of MHB type. In the central panels only
  the disc component is shown, while in the right ones both the
  material from the disc and the classical bulge is shown. The
  distance between two large tick marks is 2 initial disc
  scale-lengths. The type of the simulation is given in the upper left
  corner of the upper panels. 
} 
\label{fig:3simul}
\end{center}
\end{figure*}

The morphology of the boxy/peanut features growing in $N$-body
simulations is the same as that of boxy/peanut bulges in edge-on
galaxies. Examples of the former are shown in 
Fig.~\ref{fig:3simul}\footnote{In all the following the $z$ axis is perpendicular to the
equatorial plane and the major axis of the bar is along the $x$
axis.}. The simulation illustrated in the left panels is of MD 
type\footnote{In this example the initial conditions are
  as in A03. The specific values used for the mass, scale-length, 
  scale-height and temperature of the disc are $M_d$ = 1, $h$ = 1,
  $z_0$ = 0.2 and $Q$ = 1.2, 
  respectively. For the halo, the mass and scale-lengths are $M_h$ = 5,
  $\gamma$ = 5 and $r_c$ = 10, respectively. There is no initial
  classical bulge 
  component.}. Its bar shows a clear boxy feature
when viewed side-on and its morphology is the same as that of boxy bulges
seen edge-on. The second example is a simulation of MHB type\footnote{In this
  example the initial conditions are 
  as in A03. The specific values used for the mass, scalelength, 
  scale-height and temperature of the disc are $M_d$ = 1, $h$ = 1,
  $z_0$ = 0.2 and $Q$ = 1, 
  respectively. For the halo the mass and scale-lengths are $M_h$ = 5,
  $\gamma$ = 0.5 and $r_c$ = 10, respectively. In this simulation
  there is also a 
  classical bulge component with $M_b$ = 0.4 and $a$ = 0.4.}. Its
disc component is 
illustrated in the central column of panels. It has a strong bar and
shows a strong peanut, or X-like feature when viewed side-on.
Examining by eye a large number of simulations, shows clearly a
progression from boxy features to peanuts and then to X-features as the
strength of the bar increases. Thus, X-features are naturally linked to
strong bars seen side-on in MH type models. It is worth noting that,
when viewed from $45^{\circ}$ (third row of panels) the X-like feature
is considerably less strong and is in fact closer to a
peanut. Although in real galaxies there are 
undeniably more peanuts observed than X-features, both morphologies 
exist. X-features, some of them particularly strong, can be seen e.g. in
IC 4767, AM 1025-401, NGC 128, NGC 4845, NGC 1380A, ESO 185-G53, NGC
6771, IC 3370. 

The particular simulation in the central and right panels has initially three
components : a disc, a halo and a classical bulge. It should, however,
be stressed that the existence of an X-feature does not necessitate the
existence of the bulge component. Such features come naturally in all
simulations which develop strong bars. A02 and A03 showed that this is
easily achieved if the disc is immersed in a spheroidal component with
resonances that can absorb considerable amounts of angular momentum.
Such a spheroidal component should have considerable mass in the
resonant regions and this mass should be receptive, i.e. relatively
cool (A03). A bulge thus will help, but a concentrated halo can be
sufficient, as shown in AM02. I have,
nevertheless, decided to illustrate the X feature with a
simulation having a classical bulge, as this allows to visualise the effect
of the classical bulge. Indeed, in real galaxies it is all the
luminous material that is observed, while in
simulations it is usually only the disc component that is displayed,
in order to better illustrate its evolution. To allow for a better
comparison, I give in the right panels of
Fig.~\ref{fig:3simul} both the disc and the classical bulge component.  
If a simulation has a classical bulge component, and if it
is displayed together with the disc material, then the dip 
of the X feature at small cylindrical radii is filled in, even on
side-on views, and the simulation
displays a more boxy feature, as can be seen by comparing the
central and right columns 
in Fig.~\ref{fig:3simul}. Since the strongest bars are observed in early type
galaxies (e.g. Ohta 1996) and such
types also have classical bulges, a fair fraction of the strongest X
shapes should be observed as boxes (or perhaps peanuts). This,
together with viewing angle considerations, should
explain the relative paucity of strong X-features observed.

\begin{figure*}
\begin{center}
\setlength{\unitlength}{2cm} 
\includegraphics[scale=1.0,angle=0]{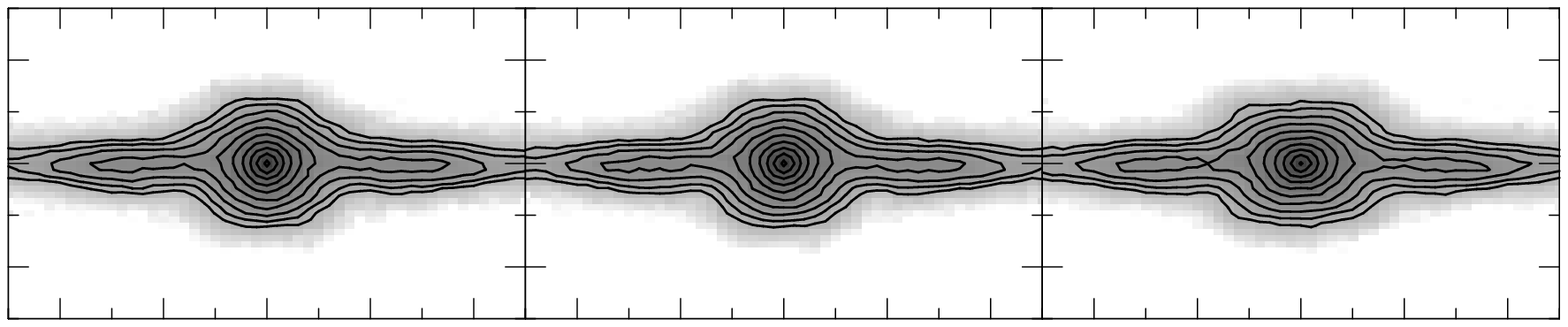}
\caption{The disc of model MHB with the bar viewed end-on (left
  panel), and with a line of sight at 5$^{\circ}$
  (central panel) and 10$^{\circ}$ (right panel) from the bar major axis. The
  distance between two large tick marks is 2 initial disc
  scale-lengths. 
} 
\label{fig:rotate}
\end{center}
\end{figure*}

Fig.~\ref{fig:3simul} also gives the end-on
views. It is important to note that bars seen fully end-on look like
classical bulges and could be mistaken for such. Thus some components
considered as bulges in edge-on galaxies could in fact be end-on
bars. Their numbers, however, should not be very large, since a
relatively small change of orientation changes the spheroidal bulge
into a boxy feature. This is shown in Fig.~\ref{fig:rotate}, which
shows the disc (only) of model MHB seen edge-on, with its bar viewed
end-on and at 5$^{\circ}$ and 10$^{\circ}$ from 
that. It is clear that, although the bar at 5$^{\circ}$ could still
easily be mistaken for a classical bulge, at 10$^{\circ}$ it could
not. 

A cursory study of the peanut strength\footnote{See AM02 for methods
  of quantifying the peanut strength} in the bars in my simulation
sample reveals further differences between the MH and MD type models. In
MD type models the peanut strength grows fast, but reaches relatively
low values. On the other hand, in MH type models the peanut strength,
like the bar strength, grows much slower, but reaches much higher
values. Further description of these differences, together with a
corresponding analysis, will be given elsewhere.
 
\subsection{Photometry : Horizontal cuts}
\label{subsec:hcuts}

\begin{figure*}
\includegraphics{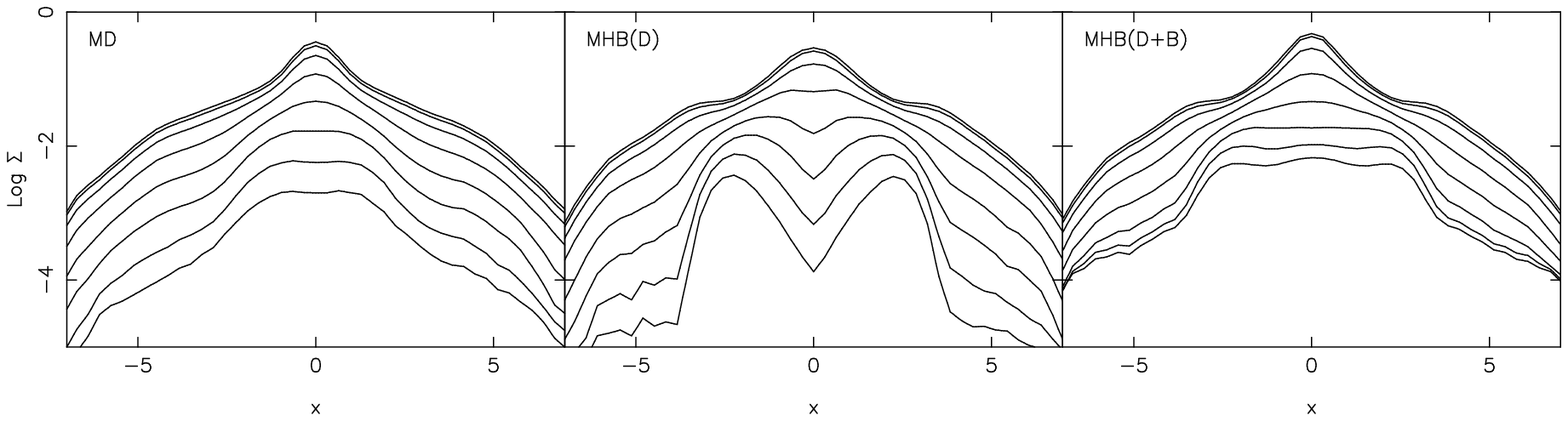} 
\vspace{5cm}
\caption{Projected surface density along cuts parallel to the major
  axis, for a side-on orientation of the bar. In each panel the uppermost curve
  corresponds to a cut at $z$ = 0 and the lower-most one to a cut at
  $z$ = 1.4. The cuts are equally spaced in $z$ with $\Delta z$ =
  0.2. The left panel corresponds to the simulation
  shown in the left panels of Fig.~\ref{fig:3simul}.
  The right and central panels correspond to the
  simulation shown in the right and central panels of
  Fig.~\ref{fig:3simul}. The projected density in the central panel was
  calculated from disc particles only, while that in the right
  panel includes the particles in the classical bulge as well. 
}
\label{fig:hcuts}
\end{figure*}

A number of photometrical studies of edge-on galaxies with boxy/peanut
bulges have obtained the projected luminosity along horizontal cuts,
i.e. cuts either on the equatorial plane, or parallel to and offset from
it. Good examples can be found 
e.g. in L\"uticke \tal (2000). Such cuts, now for the
simulations shown in Fig.~\ref{fig:3simul}, are displayed in
Fig.~\ref{fig:hcuts}. Compared to the cuts of e.g. NGC 2654 (L\"uticke
\tal 2000), or of other similarly viewed such galaxies, they show
striking  similarities. There is, nevertheless, a quantitative  
difference, namely that the  central peak in the cuts near
the equatorial plane is much stronger in NGC 2654 than in the simulation.
Simulations with a classical bulge show a
bigger peak (right panel of Fig.~\ref{fig:hcuts}), but, at least in my
simulations, not as big as that observed e.g. in NGC 2654. The latter
is presumably enhanced by stars formed from the gaseous material that
was pushed by the bar towards the central regions, and is not present
in the $N$-body simulations, since these are dissipationless. 

The remaining features are
found both in the observed and the simulated profiles. On the
profiles near the equatorial plane and outside the
central peak there is a flat ledge followed by a steep drop. This is
the signature of the bar and will be discussed further in
section~\ref{sec:objections}. The cuts at large distances from the equatorial
plane have a different aspect from those on or near the
plane. For
such cuts also there is a good correspondence between observations and
simulations. Particularly in both there is a central plateau 
whose extent is {\it shorter} than the extent of the flat ledges
in the near-equatorial cuts. The importance of this difference in
lengths will become clear in sect.~\ref{subsec:length}. On either side of
this plateau there are steep drops. For models such as MHB, if one
uses exclusively 
the disc component, or if the classical bulge is small, then cuts high
above (or below) the equatorial plane display a characteristic central
minimum. This should be also present in corresponding cuts of real
galaxies with strong X features.
If, however, one includes in simulations both 
disc and classical bulge material, then this dip fills up and a plateau can
appear instead. This is clearly seen by comparing the
central and right panels of Fig.~\ref{fig:hcuts} and should also occur
in real galaxies with both an X and a sizeable classical bulge. 

The cuts near the equatorial plane in the central and
right panels of Fig.~\ref{fig:hcuts} 
also suggest that a non-negligible part of the light that is assigned
to the bulge in observations may, in reality, belong to the old disc
component and be brought there by the mass rearrangement that
accompanies bar formation and evolution. Thus the observed central
component that constitutes the bulge according to the second definition
(see section~\ref{sec:intro}) can be 
partly due to old disc material that was pushed inwards by a bar, even
in a purely collisionless system.

\subsection{Photometry : Vertical cuts}

\begin{figure}
\begin{center}
\setlength{\unitlength}{2cm} 
\includegraphics[scale=1.0,angle=0]{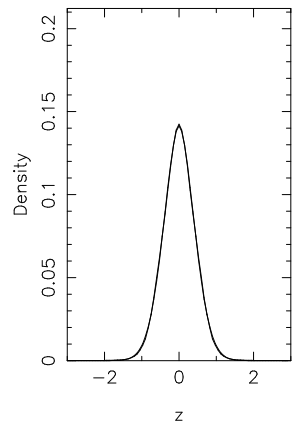}
\caption{Projected surface density along a vertical cut, i.e. a cut
  perpendicular to the equatorial plane, and, superposed, the best fitting
  generalised Gaussian (eq.~(\ref{eq:gengaus})). The two lines can 
  hardly be distinguished. 
} 
\label{fig:vcut}
\end{center}
\end{figure}

\begin{figure}
\begin{center}
\setlength{\unitlength}{1cm} 
\includegraphics[scale=0.4,angle=0]{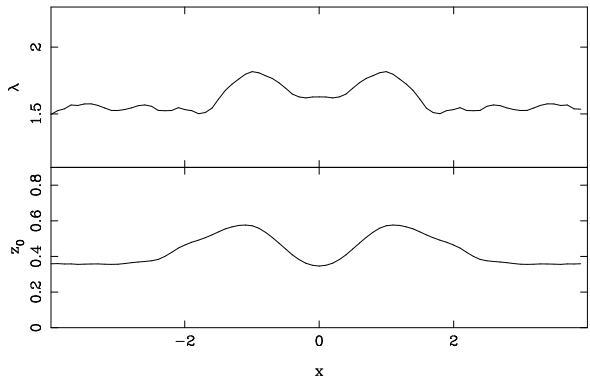}
\caption{Parameters $z_0$ (lower panel) and $\lambda$ (upper panel) of
  the generalised Gaussian as a function of the distance from the
  center of the galaxy. The fits are obtained from vertical cuts as
  that shown in Fig.~\ref{fig:vcut}. 
} 
\label{fig:z0l}
\end{center}
\end{figure}

In this subsection I will use information obtained from 
vertical cuts, i.e. cuts perpendicular to the equatorial 
plane. For this, the simulation is first viewed side-on. In order to
increase the signal to noise ratio, and since the chosen
simulations are sufficiently evolved for the peanut shape to become
symmetrical with respect to the equatorial plane, I symmetrise
with respect to the equatorial plane (up/down) and with respect to the minor
axis (right/left). I then take vertical cuts at different distances for the
center. Both for simulations and for observations, the form of the
projected surface density along the cut (i.e. $\Sigma (z)$) is
bell-like and an example from a 
simulation is given in Fig.~\ref{fig:vcut}. Such shapes can be well
fitted by generalised Gaussians, 

\begin{equation}
\Sigma (z) = \Sigma_0~exp [- (z/z_0)^\lambda].
\label{eq:gengaus}
\end{equation}

\noindent
The three free parameters, namely $\Sigma_0$, $z_0$ and $\lambda$, 
are necessary in order to fit the height, width and shape of
the density along the cut, respectively. AM02 used them to fit the log
of the (vertical) projected surface 
density, while here, as well as in Athanassoula, Aronica \& Bureau (in
preparation) and Bureau \tal (2004), they are used to fit the
projected surface density.  In this case, the generalised Gaussian 
is the same as the S\'ersic law (S\'ersic 1968), taking $\lambda$
= $n^{-1}$. These fits can be excellent. For example, Fig.~\ref{fig:vcut}
shows both the projected density along a simulated cut and the best
fitting generalised Gaussian. The fit is so good that it is hard to
distinguish two lines. This is true for the majority of the cases.

I performed such cuts and fits at a number of distances from the
center and for a
large number of simulations. Although the numerical values of the
fitting parameters and the
detailed shapes differ from one simulation to another, there are some
general trends. Results for NGC 128 are given in Fig. 3
of Bureau \tal (2004), while a typical example, from the simulation
illustrated in the left panels of Fig~\ref{fig:3simul}
can be seen in Fig.~\ref{fig:z0l}. These figures display the values of  
$z_0$ and $\lambda$ of the best fitting generalised Gaussian as a
function of the distance of the vertical
cut from the center of the galaxy. This, and similar
plots for other observations and simulations, show that
both quantities have a minimum at the center (sometimes very broad)
and a clear maxima on 
either side of it. This is true for both observations and
simulations. Furthermore, for
$\lambda$, where a direct comparison is easy since no scaling is
required, the actual values also cover 
roughly the same range in observations and simulations.       

\subsection{Median filtering}

\begin{figure}
\setlength{\unitlength}{2cm}
\includegraphics[scale=0.45,angle=0]{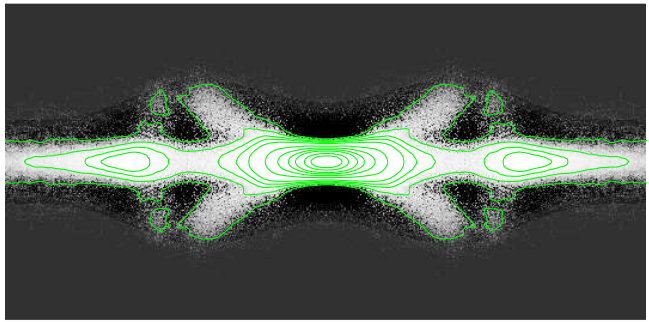}
\vskip 30pt
\includegraphics[scale=0.45,angle=0]{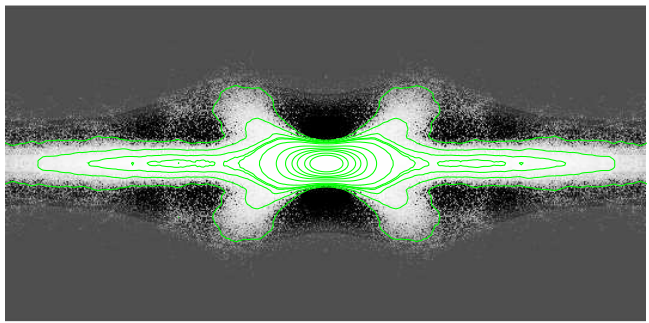}
\vskip 30pt 
\includegraphics[scale=0.45,angle=0]{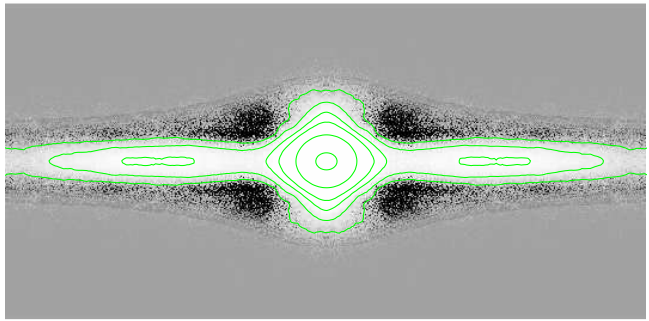}
\caption{Median filtered images for a simulation with a strong bar. In
  all three panels the disc is seen edge-on. In the upper panel the
  line-of-sight is at $90^{\circ}$ to the bar major axis (i.e. the bar is
  viewed side-on). In the middle panel it is at $45^{\circ}$ and in the
  lower one at $0^{\circ}$. Each panel includes a grey scale plot
  (with higher values having lighter shades) and some isodensities,
  chosen so as to show best certain features. The darkest areas
  correspond to negative values.} 
\label{fig:unsharp}
\end{figure}

Comparisons between observations and simulations after comparable
median filtering\footnote{This
is obtained by replacing the value of each pixel by the difference
between it and the median within a circular aperture centered on the
pixel. This highlights sharp features. It can lead to negative values,
particularly next to and just outside high density components.} is
much more demanding than a comparison of global 
morphology. This is due to the fact that median filtering reveals a
number of features, whose position and form have to be matched at a
given scale.

Aronica \tal (2003, 2004) performed median filtering of a sample of
edge-on boxy and peanut galaxies and found a number of interesting
features. In many cases (e.g. Fig. 2a in Aronica \tal
2003) one can note four extensions out of the 
equatorial plane, which form an X-like shape, except that the four
extensions do not necessarily cross the center. Another common feature
is maxima of the density along the equatorial plane, away from the
center and diametrically opposite. 
Namely, starting from the center of 
the galaxy and going outwards along the equatorial plane, the projected surface
density drops and then increases again to reach a local maximum. It
then drops to the edge of the disc. 

In order to compare the simulations to the observations, I
applied to the former a similar analysis, using the same
software. Reliable median filtering, however, requires a very large number of
particles in the disc, considerably larger than what is available in
most simulations. To remedy this I proceeded as follows. I
considered 10 snapshots closely spaced in time, so that the bar will
not have evolved noticeably, except of course for the rotation. In
practice, I found that $\Delta t$ = 0.5 in the computer units of AM02
was an adequate choice. I then 
rotated these frames so that the major axis of the bar coincides and
stacked them. Using also the natural four-fold symmetry of the problem, this
brings the number of disc particles to a total of 8 $\times 10^6$ particles. I
then viewed the disc edge-on and from the projected
surface density of the disc material I produced  
a fits image using NEMO (Teuben 1995). I then performed median filtering 
in IRAF, choosing a circular aperture of
radius one initial disc scale-length. The results for three
different viewing angles are shown in Fig.~\ref{fig:unsharp} for a
simulation with a strong bar. The
similarity between the features in the median filtered 
observations and simulations is stunning.

In the upper panel of Fig.~\ref{fig:unsharp} the bar is viewed
side-on. It displays a clear X-like form, as do the observed peanut
galaxies. It also displays secondary maxima on the equatorial plane
on either side of the center. It further shows two very faint features, like
parentheses enclosing the X.
The middle and lower panels of Fig.~\ref{fig:unsharp} 
show the same simulation but from $45^{\circ}$ and $0^{\circ}$ viewing
angle,
respectively. Viewed side-on, the four branches of the X do not cross
the center. This is probably still true, but less easy to see, when the
viewing angle is $45^{\circ}$. Furthermore, the outermost isodensity
contours joining the two upper (or lower) branches of the X look
curved. This can also be seen in a number of the median filtered 
galaxy images in Aronica \tal (2004), which might mean that
these galaxies are seen from viewing angles similar to
$45^{\circ}$. Finally, the secondary 
maxima along the equatorial plane can be seen from all three viewing
angles, but are best in the side-on view. All
these features were seen in the median filtered images of the galaxies
in Aronica \tal (2004). They are not accidental; they correspond to
specific structures of the periodic orbits that constitute the
backbone of barred galaxies, i.e. the orbits of the $x_1$ tree (SPAa),
or, more specifically, the members of the x1v1, x1v4 and z3.1s families
(PSA02). A more
thorough comparison of the observations, the simulations and the
periodic orbit structure will be given elsewhere.

\subsection{Cylindrical rotation}

For a number of galaxies with peanut/boxy shaped features there exist
published stellar velocity 
data and these show clearly that the velocity depends only
little on the distance from the equatorial plane. Good examples are 
NGC 4565 (Kormendy \& Illingworth 1982), IC 3379 (Jarvis 1987),
NGC 3079 (Shaw, Wilkinson \& Carter 1993), NGC 128 (D'Onofrio 1999)
and NGC 7332 (Falc\'on-Barroso \tal 2004). 
Such a rotation is often referred to as cylindrical rotation. 
$N$-body bars, observed edge-on, show a similar velocity structure (CDFP90;
AM02). A good example is shown in Fig.~12
of AM02, which also shows that for stronger bars the cylindrical rotation is
clearer and concerns a larger fraction of the bulge/peanut feature.
 
\subsection{PVD diagrams : gas kinematics}

Emission line spectroscopy of boxy/peanut galaxies (Kuijken \&
Merrifield 1995; Merrifield \& Kuijken 1999; Bureau \& Freeman 1999)
showed that their major 
axis position velocity diagrams (hereafter PVDs) show a number of interesting
features. Their connection to bar signatures was first made by
Kuijken \& Merrifield (1995). Bureau \& Athanassoula (1999) superposed
periodic orbits in a 
standard barred galaxy potential to study these PVDs. Their
results stressed that gaps between the signatures of the different
orbit families, as well as material in the so-called forbidden
quadrants, are a direct result of the superposition of the various
periodic orbit families. Athanassoula \& Bureau (1999) used the gas flow
simulations of Athanassoula (1992) viewed edge-on to model such PVDs. 
Shocks along the leading edges of the bar and the corresponding inflow
lead to a characteristic gap in the PVDs, between the signature of the
nuclear spiral (whenever existent) and the signature of the
disc. There is in general very good agreement between the signatures
of the observed emission line PVDs and those obtained by hydrodynamic
models. 

\subsection{PVD diagrams : stellar kinematics}

Chung \& Bureau (2004) presented long-slit absorption line kinematic
observations of 30 edge-on disc galaxies, most with a boxy peanut
feature, while Bureau \& Athanassoula (2005) `observed' in a similar
way many $N$-body bars seen edge-on. The two studies used  
for their analysis exactly the same techniques and, to where possible,
also the same software. Both used 
Gauss-Hermite series and produced profiles of the integrated
light, the mean stellar velocity $V$, the velocity dispersion
$\sigma$, as well as the higher order components $h_3$ and $h_4$. It is thus
particularly straightforward and meaningful to make comparisons of the
results of the two studies.

The similarities are striking. Both studies found the same
characteristic signatures, one in the observations and the other in
the simulations. The 
integrated light along the slit (equivalent to a major-axis light
profile) has a quasi-exponential central peak and a plateau at
intermediate radii, followed by a steep drop. The rotation curve has a
characteristic double hump. The velocity dispersion has a  
central peak, which in the center-most part may be rather flat or may
even have a central minimum. At intermediate radii there can be a plateau
which sometimes ends on either side with a shallow maximum before
dropping steeply 
at larger radii. $h_3$ correlates with $V$ over most of the bar
length, contrary to what is expected for a fast rotating disc. All
these features are spatially correlated and are seen, more 
or less strongly, both in the 
observations and in the simulations. 
The $N$-body simulations show clearly that the strength of these 
features depends on the strength of the bar. Furthermore, those features
can be interpreted with the help of the orbital structure in barred
discs (Bureau \& Athanassoula 1999).

The only observed feature that was not found by Bureau \&
Athanassoula (2005) is the anti-correlation of $h_3$ and $V$ in the
center-most parts. Indeed, the observations show that, in a small
region very near the
center, the $h_3$ and $V$ curves anti-correlate in many cases,
while for the 
remaining bar region they correlate. On the hand, in the
simulations shown by Bureau \& Athanassoula (2005) the $h_3$ and $V$
curves always correlate all the way to the center. This small discrepancy can
be remedied in two ways, presented and discussed elsewhere (Athanassoula,
in prep.). The first relies on the existence of a gaseous inner
disc, present in the galaxies but not in the simulations, while the
second relies on a different type of halo 
profile, and in particular a more centrally concentrated one. Thus
even this small discrepancy between observations and simulations can
be remedied.

\section{objections}
\label{sec:objections}

In the previous section I reviewed a number of arguments in favour of
boxy/peanut bulges being simply parts of bars viewed edge-on. This
point of view, however, is
not yet generally accepted. For example Kormendy (1993) qualifies it
as `extreme' and presents a number of arguments against it (see
also Kormendy \& Kennicutt 2004 for a more extensive discussion of
these arguments). I answer these objections here.

\subsection{Length of peanuts versus length of bars}
\label{subsec:length}

A first objection concerns the length of peanuts and bars. Indeed, ``if
boxy bulges are edge-on bars, then the longest major axis of boxy
bulges should be equal to the length of bars in face-on galaxies''
(Kormendy 1993, p. 223). This should indeed 
be the case if the {\it whole} of the bar was the
peanut. Orbital structure studies, however, have shown that this is
clearly not the case. Pfenniger (1984), in the first careful study
of 3D orbital structure in barred potentials, showed that there are
several families of 3D orbits, bifurcated at the vertical
instabilities of the main planar family, constituting the backbone of
the bar. This work was supplemented and extended in SPAa and SPAb,
some results of which I will briefly recall here, since
they fully answer this particular objection.

2D orbital structure studies clearly established that it is the stable
members of the $x_1$ family of periodic orbits that are the backbone
of a 2D bar. (Contopoulos \& Papayannopoulos 1980; Athanassoula \tal
1983; etc). Such orbits trap around them regular, non-periodic orbits
and, due to their appropriate size, shape and orientation, can form
the bar. 

\begin{figure}
\includegraphics{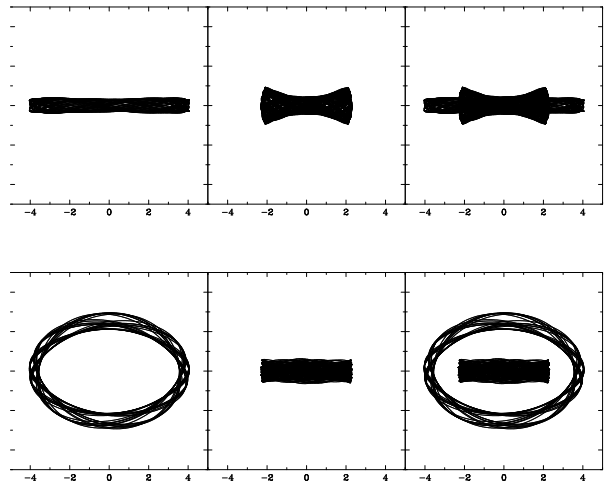} 
\vspace{7cm}
\caption{Face-on (lower panels) and side-on (upper panels) views of
  two orbits in a barred 
  potential, originating from a simulation with a strong bar and
  peanut. Both orbits are trapped around periodic orbits which are
  members of the $x_1$ 
  tree. The right panels show the superposition of the two orbits. 
}
\label{fig:2orbs}
\end{figure}

3D orbital structure studies, however, showed that the situation is
considerably more complicated. The backbone of 3D
bars is the $x_1$ tree, i.e. the $x_1$ family plus a tree of 2D and 3D
families bifurcating from it (SPAa; SPAb). Each of these families has
its own extent along the bar major axis and its own height above the
equatorial plane. Since the extent of the box/peanut will in general
be determined  by a 
different family from that which determines the length of the bar, it is
natural for the peanut and bar to have different extents in the same
galaxy. The length of the bar
is determined by the family with the largest extent in the 
direction of the bar major axis. This is usually the $x_1$ family or
2D rectangular-like orbits at the radial 4:1 resonance region (PSA03).
On the other hand, the length of the
peanut is determined by the extent of the 3D family that constitutes
its backbone, e.g. the x1v1, x1v4, etc. All this is discussed in detail
by PSA02, who also give tables with
values of the ratio of bar to peanut extent for the most important
families. It is also illustrated in Fig.~\ref{fig:2orbs}, where I plot two
orbits taken from an $N$-body simulation, as well as their
superposition. The orbit on the left is trapped around a periodic
orbit of the type defining the bar length. The orbit in the central
column of panels
is trapped around a periodic orbit of the type defining the peanut.
Their superposition (right panels) shows clearly that the
extent of the peanut can be considerably shorter than the extent of the bar. 

The fact that the bar and the peanut have different radial extents is
also clearly seen in $N$-body simulations. Examples can be seen in
Fig.~\ref{fig:3simul}. Comparison of the face-on and side-on
views (first and second row of panels) shows clearly the 
difference in peanut and bar lengths, the former being 
shorter than the latter. Several other examples are shown
e.g. in CDFP90, AM02, A02, A03, etc. The difference between the two
radial extents can also be seen clearly in Fig.~\ref{fig:hcuts}, since
the extent of the central plateau is shorter than the extent of
the flat ledges in the near-equatorial cuts. AM02 discussed and
assessed several methods by which the extent of the peanut and of the 
bar can be measured and thus quantified their ratio.

The above show that both 3D orbital structure studies and $N$-body
simulations clearly explain why the length of the peanut is 
shorter than that of the bar. Once this is understood, the difference
in extents is no more an objection to
boxy/peanut shapes being edge-on bars. On the contrary, it becomes an
argument in favour, since it is in agreement with theoretical
predictions. A stricter terminology may, nevertheless, be called for,
since strictly speaking the peanut is not the
whole of the bar seen edge-on, but only its part that sticks well out
of the equatorial plane.

\subsection{The thickness of bars, NGC 4762 and NGC 7582}
\label{sec:thickness}

Once the answer to the first objection has been understood, then the
other objections can be answered with very similar arguments. These
concern the thickness of bars, and two specific galaxies : NGC 4762 and
NGC 7582 (Kormendy 1982, 1993; Kormendy \& Kennicutt 2004).

The major-axis surface brightness profile of edge-on disc galaxies
with boxy/peanut structures  
reveals plateaus extending on either side of the center and
terminating by a more or less abrupt drop into the disc component
(e.g. L\"utticke \tal 2000). Since in more face-on systems such
plateaus are associated with bars, they have been linked with bars in
edge-on systems as well. The fact that cuts parallel to the
equatorial plane show these plateaus only if they are very little
offset in $z$   
has led to the misconception that bars
should be thin (e.g. Kormendy 1982, 1993). Before I analyse the basis of this
misconception, let me first note that such plateaus need not
necessarily be tell tale of a bar. Indeed, they could be produced by
any component which has the relative size of a bar and has abrupt
edges. 
Lenses have both these properties (e.g. Kormendy 1979, 1982)
and thus could well be responsible for a number of the observed
plateaus.

Based on the discussion in the previous subsection, which is in turn
based on a thorough study of the orbital structure in barred galaxies
(SPAa; SPAb; PSA03; PSA03),
it is easy to understand where the arguments
in favour of thin bars go wrong. The orbits that 
constitute the bar/box/peanut are, as already mentioned, part of the
$x_1$-tree. Some of the families that constitute this tree are 2D, so
that the non-periodic orbits that are trapped around their stable
members will necessarily stay near the equatorial plane. When seen
edge-on, they form the plateaus in the radial density profile. On the
other hand, some of the 3D families can reach large $z$ distances and
constitute the boxy/peanut feature. Yet both sets of 
families (2D and 3D), when seen face-on, constitute the bar
together. Thus, bars seen edge-on can have both a thin and a 
thick part, but the two can not be distinguished in face-on views. 

NGC 4762 is an object with a structure that is similar to, yet somewhat
more complicated than, the galaxies shown in L\"utticke \tal (2000).
Its photometry (Wakamatsu \& Hamabe 1984) reveals a plateau
in cuts on, or near, the equatorial plane, such as those that are associated
with bars in edge-on systems. A deep image (see e.g. Fig. 1b in
Wakamatsu \& Hamabe 1984, or the image of the SDSS) shows clearly
that it has a boxy bulge. Fisher (1997) obtained kinematical data
for this galaxy. Those show 
a number of the characteristic signatures of barred galaxies (compared
with the simulation results in Bureau \& Athanassoula 2005 and the
observations of Chung \& Bureau 2004), like the double hump in the 
velocity profile and the fact that $h_3$ and $V$
correlate over the presumed extent of the bar, except for the
innermost part where they anti-correlate. It is thus clearly a barred
galaxy seen edge-on. Seen the kinematical signatures and the shape of the
boxy feature, the bar is presumably seen nearer to end-on than to side-on.

The extra complexity of NGC 4762 comes from the fact that its radial
photometric profile has an extra
plateau, between the inner plateau and the outer disc (Wakamatsu \&
Hamabe 1984). As proposed by Wakamatsu and Hamabe, the most
plausible explanation for this structure is that NGC~4762 has both a
lens and a bar and that the bar is not seen side-on, since in the
latter case the extent of the two plateaus would be the same (Kormendy
1979). 
This argues that lenses are thin structures, but does not otherwise
affect the above arguments, according to which some of the bar
orbits when seen side-on make a clear signature (plateaus on either
side of the center) on the photometric cuts along the major axis, which
is limited only to small extents from the equatorial plane.

NGC 7582 is a particularly interesting galaxy since, despite its
intermediate inclination (approximatively $65^{\circ}$), near infrared
imaging (Quillen \tal 1997) allows us to see both the bar and the
peanut. One can now see that the peanut has a considerably shorter
extent than the bar. Similar comments can be made for NGC 4442 (Bettoni
\& Galletta 1999). As already explained above, this is not an argument
against the fact that peanuts are just edge-on bars (Kormendy \&
Kennicutt 2004), but a
confirmation of it, since it is predicted from orbital structure theory and
from simulations. 

\section{SUMMARY}
\label{sec:last}
\indent

In this paper I made a thorough comparison of the observed properties
of simulated bars viewed edge-on with those of boxy/peanut bulges in edge-on
galaxies. The comparison involves results
from the literature, as well as new simulations and, in all cases,
the same techniques are used both for the observations and the
simulations. Section~\ref{sec:comparison} includes  
a morphological comparison, properties of photometric profiles on
both horizontal cuts (i.e. parallel to the equatorial plane) and
vertical cuts (i.e. perpendicular to the equatorial plane), median
filtered images, velocity fields and PVD diagrams of both the gaseous
and the stellar component. There is excellent agreement in all
comparisons. This argues strongly that a box/peanut bulge {\it
  is simply an inherent part of a bar seen edge-on}. In order to press
this point further, 
I tackle, and answer, some previously voiced objections. I
use mainly arguments based on the 3D orbital structure in barred
galaxies, and in particular on the structure of the $x_1$ tree and of the
families composing it. In particular, I discuss the families of periodic
orbits that can be determine the extent of box/peanut bulges, and the
families that determine the bar extent. This leads to the conclusion
that the extent of the box/peanut bulge should be {\it smaller} than
that of the bar, both being measured along the bar major axis.

Boxy/peanut features are not the only types of bulges and it is now
clear that bulges are not a homogeneous class of objects. I
distinguish here three 
different types of bulges. The classical bulges, the boxy/peanut
bulges and the disc-like bulges. I briefly discuss the properties and
formation scenarios of these three types of objects and propose a
nomenclature that can help distinguish between them. I propose that
the first type of objects be called {\it classical bulges}. Members of
the second class should be called {\it box/peanut bulges},
{\it box/peanut features} (or {\it structures}), or simply
{\it peanuts}. Finally, objects in the third category could be called
{\it disc-like bulges}, since they are 
composed of disc material and have a number of properties usually
linked to discs, like their shape, radial density profile, substructure,
or kinematics. In many cases, more than one type of bulges may
well co-exist. In these cases they should be mentioned as separate
components. Thus, a given galaxy may have both a classical bulge 
and a box/peanut bulge, or a box/peanut bulge and a disc-like
bulge. In some cases all three components may be present.

The global or generic name `bulge', without adding {\it classical} or
{\it box/peanut} or {\it disc-like}, should in general be avoided. 
It is important to distinguish between these three types of objects since
they have different properties
and different formation history. It is particularly important to do this when
correlating basic `bulge' parameters with other
properties of the parent galaxy. Indeed the three different types of
bulges may follow different correlations, and thus including them all
together can, at best, add scatter, at worst, totally mask 
some relations. In other cases the relations followed by the three
types of bulges could be the same or very similar, as argued e.g. by
Kormendy \& Gebhardt (2001) for the correlation between the bulge luminosity
and the mass of the black hole it harbours. Such cases are
particularly interesting since they might be outlining physics in
which only the central concentration of the mass matters and not its
other properties. 

\parindent=0pt
\def\rr{\par\noindent\parshape=2 0cm 8cm 1cm 7cm}
\vskip 0.7cm plus .5cm minus .5cm

{\Large \bf Acknowledgments.} I thank A. Bosma, M. Bureau,
K. C. Freeman, J. Kormendy, P. Patsis, G. Aronica and M. Carollo for
useful and motivating discussions. I thank Jean-Charles
Lambert for his invaluable help with the simulation software and the
administration of the runs and W. Dehnen for 
making available to me his tree code and related programs. 
I also thank IGRAP, the region PACA, the
INSU/CNRS and the University of Aix-Marseille I for funds to develop
the computing facilities used for the calculations in this paper.
\vskip 0.5cm

{\Large \bf References.}
\rr{Aguerri, J. A. L., Balcells, M., Peletier, R. F. 2001, \AAA, 367,
  428}
\rr{Andredakis, Y. C., Peletier, R. F., Balcells, M. 1995, \MN, 275, 874}
\rr{Aronica, G., Athanassoula, E., Bureau, M., Bosma, A., Dettmar, R.-J., 
Vergani, D., Pohlen M. 2003, \ApSS, 284, 753}
\rr{Aronica, G., Bureau, M., Athanassoula, E., Dettmar, R.-J., Bosma, A., 
Freeman, K. C. 2004, \MN, to be submitted}
\rr{Athanassoula, E. 1992, \MN, 259, 345}
\rr{Athanassoula, E. 1999, in ``Astrophysical Discs'', eds. J. A. 
  Sellwood and J. Goodman,  PASP conference series, 160, 351}
\rr{Athanassoula, E. 2002, \ApJ, 569, L83 (A02)}
\rr{Athanassoula, E. 2003, \MN, 341, 1179 (A03)}
\rr{Athanassoula, E., Bienaym\'{e}, O., Martinet, L., Pfenniger,
D. 1983, \AAA, 127, 349}
\rr{Athanassoula, E., Bureau, M. 1999, \ApJ, 522, 699}
\rr{Athanassoula, E., Dehnen, W., Lambert, J. C. 2003, Highlights of
  astronomy, 13, ed. O. Engvold, 355}
\rr{Athanassoula, E., Misiriotis, A. 2002, \MN, 330, 35 (AM02)}
\rr{Athanassoula, E., Sellwood, J. A. 1986, \MN, 221, 213}
\rr{Balcells, M. 2003, in EAS Publ. ser., 10, ``Galactic and Stellar
  Dynamics'', eds. C. M. Boily, P. Patsis, S. Portegies Zwart,
  R. Spurzem and C. Theis, 23.} 
\rr{Berentzen, I., Heller, C. H., Shlosman I., Fricke, K. J. 1998,
\MN, 300, 49}
\rr{Bettoni, D. Galletta, G. 1999, \AAA, 281, 1}
\rr{Binney, J., Petrou, M. 1985, \MN, 214, 449}
\rr{Bournaud, F., Combes, F. 2002, \AAA, 392, 83}
\rr{Bureau, M., Athanassoula, E. 1999, \ApJ, 522, 686}
\rr{Bureau, M., Athanassoula, E. 2005, \ApJ, submitted}
\rr{Bureau, M., Athanassoula, E., Chung, A., Aronica, G. 2004, in
  ``Penetrating Bars through Masks of Cosmic Dust: The Hubble
  Tuning Fork Strikes a New Note'', eds. D. Block, K. C. Freeman,
  I. Puerari, R. Groess and L. Block, Kluwer Pub., 
  in press}
\rr{Bureau, M., Freeman, K. C. 1999, \AJ, 118, 126}
\rr{Carollo, C. M., Stiavelli, M. 1998, \AJ, 115, 2306}
\rr{Carollo, C. M., Stiavelli, M., Mack, J. 1998, \AJ, 116, 68}
\rr{Carollo, C. M., Ferguson, H. C., Wyse, R. F. G. 1999, ``The
  Formation of Galactic Bulges'', Cambridge Univ. Press, Cambridge}
\rr{Carollo, C. M., Stiavelli, M., de Zeeuw, T., Seigar, M., Dejonghe,
  H. 2001, \ApJ, 546, 216}
\rr{Chung, A., Bureau, M. 2004, \AJ, 127, 3192}
\rr{Combes, F., Sanders, R. H. 1981, \AAA, 96, 164}
\rr{Combes, F., Debbasch, F., Friedli, D. \& Pfenniger, D. 1990, \AAA,
  233, 82 (CDFP90)}
\rr{Contopoulos G., Papayannopoulos, T. 1980, \AAA, 92, 33}
\rr{Courteau, S., de Jong, R. S., Broeils, A. H. 1996, \ApJ, 457, L73}
\rr{Cowie, L. L., Hu, E. M., Songaila, A. 1995, \AJ, 110, 1576}
\rr{Davies, R. L., Efstathiou, G., Fall, S. M., Illingworth, G.,
  Schechter, P. L. 1983, \ApJ, 266, 41}
\rr{Davies, R. L. Illingworth, G. 1983, \ApJ, 266, 516}
\rr{Dehnen, W. 2000, \ApJ, 536, L39}
\rr{Dehnen, W. 2002, J. Comp. Phys., 179, 27}
\rr{Dejonghe, H., Habing H. J. 1993,  ``Galactic Bulges'', 
  Kluwer Academic Publ., IAU Symposium 153}
\rr{D'Onofrio, M., Capaccioli, M., Merluzzi, P., Zaggia, S.,
  Boulesteix, J. 1999, \AAS, 134, 437}
\rr{Elmegreen, B. G., Elmegreen, D. M., Hirst, A. C. 2004a, \ApJ, 604, L21}
\rr{Elmegreen, D. M., Elmegreen, B. G., Sheets, C. M. 2004b, \ApJ, 603, 74}
\rr{Falc\'on-Barroso, J. \tal 2004, \MN, 350, 35}
\rr{Fisher, D. 1997, \AJ, 113, 950}
\rr{Franx, M. 1993, in ``Galactic Bulges'', eds. H. Dejonghe and
  H. J. Habing, Kluwer Academic Publ., IAU Symposium 153, 243}
\rr{Friedli, D. 1994, in ``Mass Transfer Induced Activity in
  Galaxies'', ed. I. Shlosman, Cambridge Univ. Press, 268}
\rr{Friedli, D., Benz, W. 1993, \AAA, 268, 65}
\rr{Fu, Y. N., Huang, J. H., Deng, Z. G. 2003, \MN, 339, 442}
\rr{Heller, C. H., Shlosman, I. 1994, \ApJ, 424, 84}
\rr{Hasan, H., Pfenniger, D., Norman, C. 1993, \ApJ,, 409, 91}
\rr{Hozumi, S., Hernquist, L. 1998, astro-ph/9806002}
\rr{Hozumi, S., Hernquist, L. 1999, in ``Galaxy Dynamics'',
  eds. D. Merritt, J. A. Sellwood and M. Valluri, PASP Conference
  Series, 182, 259}
\rr{Illingworth, G. 1983, in ``Internal Kinematics and Dynamics of
  Galaxies'', ed. E. Athanassoula, IAU Symp. 100, Reidel publ., 257}
\rr{Immeli, A., Samland, M., Gerhard, O., Westera, P. 2004a, \AAA, 413, 547}
\rr{Immeli, A., Samland, M., Westera, P., Gerhard, O. 2004b, \ApJ,
  611, 20}
\rr{Jarvis, B. 1987, \AJ, 94, 30}
\rr{Kawai, A., Fukushige, T., Makino, J., \& Taiji, M. 2000,
  \PASJ, 52, 659}
\rr{Kent, S. M. 1986, \AJ, 93, 1301}
\rr{Kormendy, J. 1979, \ApJ, 227, 714}
\rr{Kormendy, J. 1982, in `Morphology and Dynamics of Galaxies',
  eds. L. Martinet and M. Mayor, Geneva Obs. Publ., Geneva, 113}
\rr{Kormendy, J. 1993, in ``Galactic Bulges'', eds. H. Dejonghe and
  H. J. Habing, Kluwer Academic Publ., IAU Symposium 153, 209}
\rr{Kormendy, J., Gebhardt, K. 2001, in ``20th Texas Symposium on
  Relativistic Astrophysics'', eds. J. C. Wheeler, H. Martel,
  Am. Inst. Phys. proc., 586, 363} 
\rr{Kormendy, J., Illingworth, G. 1982, \ApJ, 256, 460}
\rr{Kormendy, J., Kennicutt, R. C. 2004, \AR, 42, 603} 
\rr{Kuijken, K., Dubinski, J. 1995, \MN, 277, 1341}
\rr{Kuijken, K., Merrifield, M. R. 1995, \ApJ, 443, L13}
\rr{L\"utticke, R., Dettmar, R.-J., Pohlen, M. 2000, \AAA, 362, 435}
\rr{Martinez-Valpuesta, I., Shlosman, I. 2004, \ApJ, 613, L29}
\rr{Merrifield, M. R., Kuijken, K. 1999, \AAA, 345, L47}
\rr{Noguchi, M. 1999, \ApJ, 514, 77}
\rr{Noguchi, M. 1998, Nature, 392, 253}
\rr{Norman, C., Sellwood, J. A., Hasan, H. 1996, \ApJ, 462, 114}
\rr{Ohta, K. 1996, in ``Barred Galaxies'', eds. R. Buta,
D. Crocker and  B. Elmegreen, PASP Conference Series, 91, 37}
\rr{O'Neill, J. K., Dubinski, J. 2003, \MN, 346, 251}
\rr{Patsis, P., Skokos, Ch., Athanassoula, E. 2002, \MN, 337, 578 (PSA02)}
\rr{Patsis, P., Skokos, Ch., Athanassoula, E. 2003, \MN, 342, 69 (PSA03)}
\rr{Pfenniger, D., 1984, \AAA, 134, 373}
\rr{Pfenniger, D., 1993, in ``Galactic Bulges'', eds. H. Dejonghe and
  H. J. Habing, Kluwer Academic Publ., IAU Symposium 153, 387}
\rr{Pfenniger, D., 1999, in  ``The
  Formation of Galactic Bulges'', eds. C. M., Carollo, H. C. Ferguson,
  and R. F. Wyse, Cambridge Univ. Press, Cambridge, 95}
\rr{Prugniel, Ph., Maubon, G., Simien, F. 2001, \AAA, 366, 68}
\rr{Quillen, A. C., Kuchinski, L. E., Frogel, J. A., DePoy,
  D. L. 1997, \ApJ, 481, 179}
\rr{Raha, N., Sellwood, J. A., James, R. A., Kahn, F. D. 1991,
  Nature, 352, 411 (RSJK91)} 
\rr{Regan, M. W., Teuben, P. J. 2004, \ApJ, 600, 595}
\rr{Rowley, G. 1988, \ApJ, 331, 124}
\rr{Samland, M. \& Gerhard, O. E. 2003, \AAA, 399, 961}
\rr{Sandage, A.  1961, The Hubble Atlas of Galaxies, Carnegie
Institution of Washington, Washington DC}
\rr{S\'ersic, J. 1968, Atlas de Galaxias Australes, Obs. Astron. Cordoba}
\rr{Shaw, M., Wilkinson, A., Carter, D. 1993, \AAA, 268, 511}
\rr{Shen, J., Sellwood, J. A. 2004, \ApJ, 604, 614}
\rr{Skokos, H., Patsis, P., Athanassoula, E. 2002a,
\MN, 333, 847 (SPAa)}
\rr{Skokos, H., Patsis, P., Athanassoula, E. 2002b,
\MN, 333, 861 (SPAb)}
\rr{Sommer-Larsen, J. G\"otz, M., Portinari, L. 2003, \ApJ, 596, 47}
\rr{Steinmetz, M., M\"uller, E. 1995, \MN, 276, 549}
\rr{Steinmetz, M., Navarro, J. 2002, New Astronomy, 7, 155}
\rr{Teuben, P. J. 1995, in ``Astronomical Data Analysis Software and
  Systems IV'', eds. R. A. Shaw, H. E., Payne and J. J. E. Hayes, ASP
  Conf, Ser. 77, 398}
\rr{Wada, K., Habe, A. 1992, \MN, 258, 82}
\rr{Wada, K., Habe, A. 1995, \MN, 277, 433}
\rr{Wakamatsu, K., Hamabe, M. 1984, \ApJS, 1984, 56, 283}
\rr{Webster's New Twentieth Century Dictionary of the English
  Language 1963, second edition, The World Publishing Company,
  Cleveland and New York}
\rr{Wyse, R. F. G., Gilmore, G., Franx, M. 1997, ARA\&A, 35, 637}
\label{lastpage}

\end{document}